# Exploring the outsourcing relationship in software startups – A multiple case study

Anh Nguyen Duc
Department of Computer and Information
Science (IDI), NTNU
NO-7491 Trondheim, Norway
anhn@idi.ntnu.no

Pekka Abrahamsson
Department of Computer and Information
Science (IDI), NTNU
NO-7491 Trondheim, Norway
pekkaa@ntnu.no

***Abstract –*** Software startups are becoming increasingly popular in software industry as well as other sectors of economy. Startups that lack necessary competences often seek for external resources from outsourcing partners. Little is known how this outsourcing relationship works and whether it makes sense to outsource the technical competence to an external party. This is among the first investigations on the outsourcing relationships in software startups. By conducting exploratory case studies at six startups, we found a mixed experience with outsourcing. The experimental nature of an early product development makes outsourcing a feasible option, although startups often suffer from its uncertainty and managing commitments from partners. Results further propose that early contract-based activities could be transformed into a long-term partnership by adopting a startup boundary spanner's role, establishing an inter-personal relationship and maintaining a mutual commitment.

***Keywords: outsourcing partnership, outsourcing startup, global software development, software startups, empirical study, exploratory case study,***

## I. INTRODUCTION

Software industry has witnessed a growing trend of software products developed by small entrepreneurial teams aiming at a scalable business model [1]. A typical early stage startup is often financially bootstrapping, starting with some forms of initial self-funding (sweat equity, credit, savings, etc.) [7]. They rarely have both the necessary financial and human resources to complete the tasks required in the journey from ideation to commercialization. The situation forces them to reach out to external resources in performing engineering tasks such as design, prototyping, and manufacturing [11, 12]. These relationships could take a form of outsourcing, partnership, and later as acquisitions, and joint ventures [13].

To outsource or not is a controversial topic among startup enthusiasts [4-6]. In one hand, practitioners argue that outsourcing core competence, i.e. technical development in a technology-based startup is risky and can be illustrated as "*penny wise and pound foolish*" [5,6]. Moreover, hidden cost of outsourcing projects are known with geographical, temporal and cultural gaps among project locations. [9, 10]. On the other hand, we witness successful stories about adopting outsourcing strategies in unicorns at their early stages. Slack, the 3.8 billion USD valued startup, started their product development with an outsourced team [4]. Early versions of Skype, the video messaging platform acquired by Microsoft with 8.5 billion USD, was initially built by an outsourced team in Estonia [14]. More generally stated, outsourcing is considered as a strategy for innovation and radical changes in organizations [3]. It can be argued that these cases demonstrate for the viability of adopting outsourcing as an external sourcing strategy for software startups.

When outsourcing is predetermined or unavoidable, the awareness of common pitfalls and challenges would save entrepreneurs from serious managerial mistakes. Understanding the rise and evolution of the relationship in outsourcing arrangements is vital, because it not only helps through the operationalization of the agreement but also in the issues of dependency [16]. While Global software development (GSD) is rich with experience reports and best practices for outsourcing projects, it is biased to more established companies with software development processes in place [9, 10]. A review of empirical evidence shows only ten primary studies addressing GSD in the collaboration among small companies [9]. None of them purposefully target the contextual factors that are unique to startups.

A unique characteristic of a startup context is the experimental nature in the processes of both business and product development. Startup is searching for a scalable business model and often faces rapid changes at business idea, requirement and product levels. Product development in startups is considered as a set of opportunistic activities, which focus on providing value under constrained conditions of startups [23]. In a software startup context, the employed development team needs to be able to cope with many uncertainties and unknowns.

This work focuses on understanding entrepreneurs' experience with outsourcing as a team competence strategy. The study was performed under a large-scale multiple case study research about the state of practice in software startups. For this study, the insight was gathered from six companies that had adopted outsourcing strategies at various startup stages. Our specific research questions are:

- RQ1: What types of tasks are outsourced in software startups?
- RQ2: How do the outsourcing relationships evolve during the progression of a software startup?

The rest of the paper is structured as follows: Section II presents related literature, Section III describes the research approach, Section IV presents the findings, Section V discusses the results and Section VI concludes the paper.

## II. RELATED WORK

### A. The distinction between startups and SMEs

There is no consensus definition about what a startup is. With Lean Startup and Customer Development approaches, Steve Blank et al. refer to a startup as an "*organization formed to search for a repeatable and scalable business model*" [18]. A startup typically carries on a business model, which is repeatable and applicable to a large market volume. U.S. Small Business Administration describes a startup as a "*business that is typically technology oriented and has high growth potential*" [19]. The term *technology* can be applied to both software and hardware parts. In the scope of this work, we focused on software startups, who develop products with significant software parts.

**Table 1: Comparison between a startup and a SME**

| Elements | Startups | SME |
|---|---|---|
| **Business goal** | High growth | Stable business |
| **Financial Risk** | High risk | Low risk |
| **Organization structure** | Various from agile team to more structured organization | A structured and stable group of employees |
| **Funding** | Often seek large-scale funding from venture capitalists or angel investors, IPO | self-funded or financed from family, friends or a bank loan |
| **Product** | Unknown, often related to advanced technology | Often known, various |
| **Customer** | Unknown | Often known |

There are fundamentally differences between a startup and an SME in term of business, organization and product dimensions, as shown in Table 1. U.S. Small Business Administration describes SME as "*independently owned and operated, organized for profit, and not dominant in its field*." [19]. Startups are designed to grow fast while SMEs focus on doing stable business for years. Another difference lies on the product and market certainty. SMEs generally sell known products to known customers in known local markets. In startups, customer and product are often unknown from the beginning. A lot of changes to both customers and products would be expected during the startup journey. Rather than a formal organization, a startup is likely to be a task-oriented group. Entrepreneurs are central to the organization as a whole and they carry out most of tasks [20].

### B. Software startup life cylce

There exist several frameworks and models characterizing the evolution of a startup [21-24]. Reynolds et al. describe an entrepreneurial cycle with four phases, conception, gestation, infancy and adolescence [24]. Cooper's stage-gate model [21] consists of five stages from an idea to the product lunch, namely, (1) discovery, (2) scoping, (3) build business case, (4) development, (5) testing and evaluation. Eric Ries states that that startups will need to go through three abstract steps, problem-solution fit, product-market fit and scaling up [22].

In a previous work, we proposed a framework to analyze Software Engineering (SE) activities in software startups [2]. The demand of SE activities is observed to be varied across (1) idea phase, (2) pre startup phase, (3) startup phase and (4) scaling phase. Idea phase characterizes by business opportunities firstly identified, and refined through planning, initial idea validation and proposal writing. Pre-start-up phase is when minimal viable competence was gathered in a startup team, construction of prototypes and approaching early customers and funding organizations. Startup phase is marked by introducing formal legislation, involving more management and serving customers. The product development in startups is considered as opportunistic activities, which focus on providing value under constrained conditions of startups [23]. Scaling phase is marked with rapid growth of users, revenues, team and transferring into a stable organizational structure.

### C. SME engineering aspects of GSD

GSD becomes a part of everyday business with the use of different terms related to sourcing strategies, i.e. outsourcing, offshoring, near-shoring, far-shoring, right-shoring, best-shoring, etc. Smite et al. described four types of sourcing arrangements based on whether outsourced tasks happen in the same company (insourcing or outsourcing) and if they happen in the same country (on-shoring or offshoring) [26]. The success or failure of an outsourcing experience is viewed from a technical perspective, relating to the quality of delivered code, communication and customer satisfaction [34]

Outsourcing is considered as beneficial for small companies in both direct and indirect ways, by cost saving, proximity to market, skilled work forces, improving teamwork and processes [27]. In a distributed context, collaboration, coordination and collaboration are recognized as project success factors [28]. Practices of managing coordination, communication, team identity, trust, effective use of tools, and other human factors are reported for GSD projects with the participants of SMEs [29-32]. A study of European SMEs showed that key problems for establishing trusts are poor socialization and socio-cultural fit, increased monitoring, inconsistency and disparities in work practices, reduction of and unpredictability in communication; and a lack of face-to-face meetings, language skills, conflict handling, and cognitive-based trust [33]. Nils et al. described failed outsourcing attempts in four Scandinavian projects and showed that successful offshore software development requires a change from a cost-driven focus to an intellectual capital- driven focus [34].

### D. Business aspects of GSD

From business perspectives, outsourcing relationship is typically initiated as a contractual agreement to deliver certain tasks, products, or services in order to receive monetary payback [25]. Literature reveals several challenges with maintaining contract-based projects, such as (1) difficulties in writing a complete contract, (2) investment from one or both parties on relation specific assets, (3) rigidness of written agreements [35]. Besides, uncertainty and project risks are important barriers for contract-based outsourcing project [36]. In contrast to contract-based approaches, many outsourcing projects are established via a partnership. Software outsourcing partnership is defined as "*results of a process of transferring the responsibility of developing software for a specific business function from an employee group to a non-employee group including transfer of assets such as personnel*" [38]. Lee et al. described outsourcing partnership as a mechanism for (1) protecting relations specific assets investments and promote further investments, (2) sustaining long-term relationships, (3) and better dealing with uncertainty [37]. Sikandar Ali et al. summarized critical success factors in software outsourcing partnership, i.e. mutual understanding, trust, communication and dependency management [38]. In this study, we looked at both contractual relationship and partnership from startup viewpoint.

## III. RESEARCH APPROACH

### A. Research methodology

To explore state-of-practice startup development with a special focus on outsourcing activities, we conducted a multiple exploratory case study. According to Yin [39], a case study design is appropriate when (a) the focus of the study is to answer "*how*" and "*why*" questions; (b) there is a high influence of contextual factors on the studied phenomenon. Exploratory case studies were selected for the purpose of "*finding out what is happening, seeking new insights and generating ideas and hypotheses for new research*" [40]. A multiple case study enables the researcher to explore differences within and between cases. We selected an outsourcing project as a unit of analysis. We had cases with multiple outsourcing projects that we attempted to document and to portray them separately.

There is often difficult to identify a real startup case among other similar phenomenon, such as freelancers, SMEs or part-time startups. Therefore, we selected startups with at least two full-time members, operating for at least six months, having at least a first running prototype, and doing software development. By searching through three channels, namely (1) professional networks of papers' authors, (2) accelerators, co-working spaces and incubators at the locations of the authors, (3) startups listed in Startup Norway, we identified a contact list includes 219 startups from Norway, Finland, Italy, Germany, Netherlands, Singapore, India, China, Pakistan and Vietnam. After sending out invitation emails, we received 41 feedbacks, approximately 18.7% response rate. Outsourcing startups were selected as a subset from our startup cases. Excluding startups that are not interested in the research, or startups that do not pass our selection criteria, the final set of cases are 20 startups. There are seven startups (35%) that perform certain kinds of outsourcing activities, in which we were able to collect data on six of them. For concealment, the companies are not named in this paper, but instead referred to as CT1, CT2, CT3, CT4, CT5 and CT6. CT1 was selected as the first author has been a part of the management team of the startup with a lot of insight. CT2, CT3 were selected from exploring local incubators. CT4 was contacted via the Startup Norway network. CT5 was selected as the former CEO was a professor in the same department with the authors. This enables us to access a rich material about the case. CT6 was selected from the personal contact from the first author with also a lot of insight.

### B. Data collection and analysis

As triangulation is an important mechanism to increase the research creditability, we conducted data collection from multiple data sources and viewpoints. Methodological triangulation is implemented by using both semi-structured documents and observations, as shown in **Error! Reference source not found.**. In CT1, CT2 and CT6 we collected data from both startups and outsourcing partners, which provide a viewpoint triangulation. Business documents, such as business model canvases and business plans were exposed to the research team as a preliminary step prepared for interviews. We also looked at the company websites and tried to find other online source of information. In Company CT1, one of the co-authors was a participant of the startup, therefore many internal documents were available to the research. In Company CT6, one of the co-authors involved as an investor as well as a mentor, which allowed the access to internal documents and resources.

The collected data were a part of a large-scale multiple case study about software startups, in which different angles are discovered, i.e. evolution patterns, pivoting, validated learning and engineering activities. The interview guideline includes four parts (1) business background (2) idea visualization and prototyping (3) product development (4) challenges and lessons learnt. While the general interviews provided insights of contextual information and current states of our cases, we also conducted several in-depth follow-up interviews that focuses on outsourcing activities.

In CT1, we had several participant observations in the duration from March 2016 to November 2016. In one hand, the first author participated in every Sprint planning and retrospective meetings with the outsourced team. In the other hand, the first author also participated in business meetings with the CEO of CT1 and exposed to different stakeholders, i.e. accelerator program, investors and mentors.

In CT2, interviews with both a startup CEO and an outsourced team leader was performed. Interviews were conducted three months from each other, which enabled the reflection on the evolution of the startup. We were also allowed to access issue tracking system and code repository of the company for research purpose.

**Table 2: Data collection**

| Cases | Interview | | Documents, Observations |
|---|---|---|---|
| | CEO | Outsourced | |
| CT1 | 3 | 2 | Outsourcing agreement, product description, Sprint meeting minutes, test plan, repository, issue tracking system |
| CT2 | 1 | 2 | Website, project plan, repository |
| CT3 | 1 | 0 | Website |
| CT4 | 1 | 0 | Website, media presses |
| CT5 | 2 | 0 | Website, media presses |
| CT6 | 1 | 1 | Outsourcing contract, pitching documents, project description |
| Total | 14 | | |

To enhance findings from other cases, we added CT6 to compare and contrast the findings. In total we collected fourteen interviews from CEO, CTO and developers in the six startups.

To support the data analysis, we used a qualitative analysis tool so-called NVivo 11[2], which enabled classification and analysis of textual data and the summary of extracted codes. In the first dimension, relevant textual information was extracted and synthesized by narrative synthesis [8]. The narrative description with relevant quotes were selected, interpreted and ordered by different phases of a startup life cycle (answering the RQ2). In the second dimension, we grouped and interpreted the experience of interviewee about outsourcing as positive or negative for each case. A tailored cross-case analysis was done by comparing and contrasting the outsourcing experience regarding to technical tasks (answering the RQ1) and contract-based and partnership (answering the RQ2).

*C. Case description*

Overall description of the cases is given in Table 3. The information is described for the latest stage of the startups, i.e. current startup's product, the current startup headquarter location, and current business models. As described, our cases include startups about marketplaces, business management, game and IoT devices. The application domains include education, journalism, transport and aquaculture. Most of the startup cases initiated in Norway (expect CT6). Company size ranges from two to 17 people, at the time of the interviews. The case's foundation years vary from 2012 to 2016. In two cases, which has passed a start-up phase, they were still able to reflect on their experience in the earlier phases of their operating history.

[2] http://www.qsrinternational.com/product

Company CT1 is a spin-off from a social media cooperation. The CEO quitted her job and seek for a technical team to develop a hyper-local news platform. The CEO used different freelancers and contractors to develop and to illustrate for the business plan. After that, a CTO joined the team and CT1 started a prototyping contract with a Vietnamese outsourcing team. The team was selected after a bidding process to ensure the lowest price quote. The contract was made based on six-milestone delivery and payment was made after each milestone. The outsourcing team worked in a Sprint-based approach adopting Sprint planning and retrospective meetings, burn-down chart and communication via social media. After nine months of collaboration, the CEO stated that this is a positive experience regarding the value perceived. The outsourced team was offered to be a part of the startup.

Company CT2 was founded by a CEO with many years of consulting experience in construction business. The CEO had an idea of making a mobile solution for collaborating information across different departments of a construction project. After three months participating in an incubator program in USA, she returned to Norway and started the company. Outsourcing was selected in early 2012 to save cost and to achieve high-quality developers. The first contract was to implement the most important feature as proof-of-concept version and adopted in a customer organization. After several prototypes, the company launched their product in late 2012. At the moment, the company passed the breakeven point, had more than 5000 users and planned to extend to neighbor markets. While the CTO is currently skeptical about the high dependence on the Indian outsourcing team, the CEO expresses the satisfaction with the team. A partnership is established by giving the leader of the Indian team a part of company shares.

Company CT3 was founded in late 2012 after six months of market research. The business idea was to facilitate the event organization and purchase in Norway. The team hired an India development team from early 2013. The CEO reports a challenge in the outsourcing relationship, which leads to the termination. In 2014, the CEO did another outsourcing project to Ukraine and had a positive experience. The company had 200 customers so far with about 10.000 transactions in 2015. The company used to have five people based in Norway and an outsourcing team of eight developers. The current payroll only includes three employees in Oslo.

Company CT4 was founded in September 2013, by launching a commercial product like Airbnb for shipping services. Started from Norway, CT4 moved the headquarter to London in December 2015. For the first two years, CT4 focused on product development. The development was iterative and evolutionary. The CEO had two major projects outsourcing to Greece and Germany, which both turned out to be failure attempts. The outsourced location was selected by considering cultural, time-zone and geography matching. The CEO chose Germany later with the intention to explore German market. Currently, the company did all of their engineering activities in-house.

Company CT5 was initiated by a professor in a Norwegian university. The idea was to motivate student's learning in class by playing quizzes. Initiated from 2007, the idea was developed and refined within academia environment. In 2013, the idea was commercialized and soon after that the headquarter is moved to London, UK.

**Table 3: Software startup demographic**

| Case | Product | Started year | Biz. type | Loc. | Dev. approach | CEO's backgrd | Funding | # Ppl. | Latest Stage |
|------|---------|--------------|-----------|------|---------------|---------------|---------|--------|--------------|
| CT1 | Hyper-local news platform | 2015 | P2P | Norway | Agile | Business | Bootstrap | 6 | Startup |
| CT2 | Collaboration platform for construction | 2012 | B2B | Norway | Distributed Scrum | Business | Bootstrap | 9 | Scaling |
| CT3 | Ticket and event system | 2012 | B2B | Norway | Agile | Business | Bootstrap | 3 | Startup |
| CT4 | Shipping platform and services | 2013 | P2P | UK | Agile | Business | Early investor | 5 | Startup |
| CT5 | Game-based classroom learning tool | 2013 | P2P | UK | Distributed Agile | Technical → Business | Bootstrap | 12 | Scaling |
| CT6 | Fish farm management and tracking | 2016 | B2B | Vietnam | Adhoc. | Technical | Bootstrap | 5 | Pre-startup |

While early prototyping and product development was done by students, the later version was developed by in-house developers and sub-contractors. The product is used by more than 50 million people in 180 countries. In 2015, CT5 was invested two million USD. Currently, the development and maintenance of the product was done in both locations in UK and Norway. The team followed a distributed Agile model with remote meetings and frequent visits. The interviewee, used to be the CEO of the company, expresses the positive experience with the collaboration model.

Company CT6 is a bootstrap Vietnamese startup that develops a IoT solution for fish farm tracking and management. The startup teams composed of three software developers, one designer, one hardware developer and the CEO. The team participated in an incubator program in Korea for three months. Several low-fidelity prototypes were made during the business concept development. A hardware prototype was contracted to a local research institute. There were already early customers that invested in the product. Currently, the software part is developed in-house and the hardware production was sent to China. The CEO expresses the satisfaction with the choices of contractors due to the cost, time and quality of delivery.

### D. Threats to validity

There are some relevant considerations related to external validity, internal validity, and construct validity [1]. Case study research is often criticized for external validities as it relates to the generalization of study findings. Our cases were selected to demonstrate for different development phases of startups, and also for both successful and failure experience with outsourcing. Our sample is characterized by Norwegian software startups, with a small team and bootstrap financing model. We do not consider other types of startups, for example, internal cooperate startups, venture capital invested startups, and USA-based startups. Hence, the results cannot be directly applied to other contexts, though analytical generalization may be possible in similar contexts.

One internal threat to validity is the fact that our evidence mainly based on semi-structure interviews. In order to mitigate this threat, we selected CTO and CEO as interviewees, who have the best understanding about their startups. We also triangulate our arguments from both startups and outsourcing team's perspectives. A construct validity threat is the possible inadequate descriptions of constructs. We tried at our best to collect contextual information about the startups, from social media and personal contacts. When analyzing data, the coding process of interview transcripts was assisted by the authors' prior knowledge about prototyping and validated learning. This helped to focus on the investigated phenomenon without losing relevant details.

## IV. RESULTS

### A. RQ1: What types of tasks are outsourced in software startups?

We identify two types of activities that were outsourced in startup companies, namely engineering and non-engineering, (as described in Table 4 ).

Engineering tasks include prototyping, conceptualization, product design, and also full-scale development. Prototyping is contracted to outsourced partners in CT1, CT2, CT4 and CT6. Such outsourced tasks are small and well-defined, with the scope from few days to few weeks. In CT1 and CT4, the CEOs hired contractors to develop low-fidelity mockups that complement for their business plan. In both cases, prototyping contractors were recruited early when the CEO does not have the expertise in hand. Similarly, CT6 consisted of full-stack software developers initially. Lacking of a hardware developer in team made the CEO selected a contractor for developing a hardware part of the prototype.

**Table 4: Outsourcing tasks in startups**

| Type | Activities | Description | Case |
|------|-----------|-------------|------|
| Engineering activities | Prototyping | Low or high fidelity prototype, i.e. website, mobile apps, IoT devices | CT1, CT2, CT4, CT6 |
| | Product conceptualization | Requirement collection, influence on a feature list | CT1 |
| | UX Design | Mobile, website interface design | CT1, CT2 |
| | Implementation | Code, testing, integration for a given backlogs | CT4, CT5 |
| | Full-scale development | Implementation, testing, maintenance | CT1, CT2 |
| | Development and Operation Infrastructure | Web hosting, data storage, data-related services | CT1, CT2, CT3, CT4, CT5 |
| Non-engineering activities | Branding and marketing | Branding design, marketing plan and strategy | CT1 |
| | Accounting and Administrative | Payroll, tax calculation and report, etc | CT1, CT6 |
| | Content development | Seed content, blog, news, etc | CT1 |

Outsourcing full-scale product development occurs in CT1, and CT2. In CT1, an outsourced team also involved in business development activities, such as user's interviews and focus groups. They were given the freedom to collect user's feedback from the team's location and to contribute to the product's requirement. In both cases, the outsourced team proposed choices of technology, architectural design and implementation.

In CT1, a milestone-based test plan was written by the CTO to make sure different quality aspects were covered. User acceptance test and system testing is done from the client side. Similarly, the CEO of CT2 involved intensively in a front-end testing and user testing of the product. It is the same situation in the early time of CT3 where the CEO does not have a technical team and relies on outsourced developers. Tasks, such as requirement elicitation and front-end design were conducted with a close involvement of the client. In CT1 and CT2, product maintenance is also handled by the outsourcing teams, including fixing bugs, developing new features and modification request.

We observed in all cases the adoption of cloud-based services, for instance, Platform as a Service (Google AppEngine in CT6), Infrastructure as a Service (Amazon Web Services in CT1, CT3, CT4, Microsoft Azure in CT2). In CT1 and CT6, cloud computing offers an ability to quickly deliver the (prototypes of) services to users, simplify management and deployment tasks. The selection of services is often suggested by the outsourcing team, or requested by CTOs of startups.

Due to lack of resource, startups outsource legal and administrative tasks to a low-cost contractor. Tax calculation and report is done by local financial service providers in CT1, CT4, and CT6. Occasionally, startups also hire professional to perform marketing activities and content development for their products, as illustrated in Table 4.

### B. RQ2: How do the outsourcing relationships evolve during the progression of a software startup?

In the scope of this work, we focus on experience with outsourcing engineering tasks. We summarized the evolution of outsourcing relationships in our cases in Table 5. There are eleven outsourcing experience as CT1, CT2 and CT4 have outsourced multiple times. The experience is described in term of the phases it happened (idea, pre-startup, startup or post-startup), the nature of the relationship (contractual or partnership) and the positive (marked as green) or negative (marked as orange) experience.

#### 1) Idea and pre-startup phase

During the pre-startup phase, startups engage in contractual relationships, such as outsourcing, freelancing and contractors, to explore the problem-solution fit, by performing market research, business concept development and preliminary prototypes. A typical arrangement for non-technical founder would be to build multiple design concepts and engineering mock-ups by contracting outsourcing partners (CT1, CT2, CT3, CT4, CT6). In this phase, the contractual relationships with outsourcing partners were explicit and discrete. Startup CEOs often follow closely with terms and agreements made in contracts. The arrangement about functional and non-functional requirements were set as appendixes to the contracts (CT1). The main motivation was to achieve professional deliverables in a cost-saving manner. The CEO of CT2 mentioned that hiring a skillful developer in Norway would be too expensive and also difficult for her to know their capabilities.

> *"First I have a proof of concept with two different Indian companies to test which one I should go for….. Having the product developed outside, it can save probably [Amount] NOK for us"* (CT2)

**Table 5: Outsourcing relationship in startup's phases**

| Cases | Outs. Location | Idea &Pre Startup | Startup | Scaling |
|-------|-------|-------|-------|-------|
| CT1a | Pakistan | Contractual | N/A | N/A |
| CT1b | Norway | Contractual | N/A | N/A |
| CT1c | Vietnam | Contractual | Partnership | N/A |
| CT2a | India | Contractual | N/A | N/A |
| CT2b | India | Contractual | Contractual | Partnership |
| CT3a | India | Contractual | N/A | N/A |
| CT3b | Ukraine | Contractual | N/A | N/A |
| CT4a | Germany | Contractual | N/A | N/A |
| CT4b | Greece | N/A | Contractual | N/A |
| CT5 | UK | N/A | N/A | Acquisition |
| CT6 | China | Contractual | N/A | N/A |

As their business growing, startups learned about outsourcing partner's capacities in delivering in both quantity and quality manner. CT1 expressed a bad experience for choosing a cheap freelancing service from an online job portal, as the CEO stated her doubt about the experience and track record of the partner:

> *"The platform is [Freelancing Platform]. The designer was okie but the developers were horrible. They didn't test the website with both iphone and android platforms. I took so many iterations to get to what is acceptable."* (CT1)

Establishing a smooth coordination and communication is as critical in startup situations as it is known in GSD [28]. Due to the cost saving strategy, most of the outsourcing partners locate in different geographical, temporal and cultural environments [10]. It is often a challenge to establish an outsourcing relationship with a stranger from a freelance portal. It is also a learning experience with known partners:

> *"… It was the problems of communication with the Indian company. It appeared as a cultural barrier because instead of saying okay it took one more month then we would maybe say tomorrow or next week. So, we did not understand that and it was quite frustrating.…unforeseen costs and challenges apparently."* (CT3)

It is also problematic for startuppers who are not familiar with practices of remote collaboration, as described in CT3:

> *"The second main challenge is the fact that you work remotely is less effective than if you are all sitting in the same room."* (CT3)

In this case, the outsourcing relationship was soon terminated and startups searched for alternative solutions. In some cases, they also switched to find other source of technical competences, i.e. doing it by themselves, or asking for helps from friends.

When dealing with deliverables, startups give a freedom and responsibility to partners who they trust. For example, CT1 took inputs from a local contractor to provide a product solution and suggestions at strategic level. This is because the CEO had worked with this contractor before and knew their ability and competence. In CT6, the CEO was searching for a partner to make a mock-up for an IoT device. He trusted an R&D institute that is associated with his formal university and is operated by his supervisor before. Specifically, startups expected their trusted partners to perform more integrated and complex works. CT6 originally planned to implement many functionalities in-house. But after their contractor exceeded the expectation, they decided to increase the portion of outsourcing task, including a complex design and detailed implementation of core modules.

In summary, during the pre-startup phase, startups consider external resource from a transactional relationship to save cost and to utilize expertise that are shortage in-house. Cultural, geographical boundaries and lack of experience on remote collaboration often kill the outsourcing relationship. In some cases, outsourcing is a failure experiment when gained value is not as expected. When the startup's expectation is met, trust began to form.

*2) Startup phase*

After the product ideas being conceptualized and prototyped, requirements become known for full-scale development. Two of our cases continued the outsourcing relationship based on contractual exchange with the previous partners they had in Idea and Pre-startup phase (CT1, CT2).

> *"We decided to hire a software company in India to develop the product. We start developing what we are having today…" (CT2)*

When shifting to a new outsourcing team, we observed a negative experience with technical debt and code legacy in CT4:

> *"It's hard to get a person that didn't develop the system to do the maintenance. … having somebody that has fully ownership with [Source Code] that will work on it in the longer run is better" (CT4)*

The establishment of contract-based trust and mutual understanding was the key in expanding into a more integrated and co- dependent relationship. Startups that dissatisfied with their outsourcing partners' work performance did not continue the relationships into the startup phase (CT4). It is important in startup collaboration that dynamic and rapid changing natures of startups are understood. The good competence might not be enough when outsourcing tasks relate to core values of the startup:

"*Well, the main challenge is that they [outsourced team] are not our team… It is a smart team and they are good people that know what they are doing… What should I say, they don't care about us. They care about their own company*" (CT4)

As software startups operate under multi-influence and constraints, the product development can be formed as an exploratory journey with unexpected technical and feature changes. Outsourcing partners would be expected to take more commitment in the journey:

"*They do their work, they clock in at nine in the morning and they go home at six. It is fine but they will never love what we do or the product that we are working on, which is why we are now moving into an in-house house team. I think that is the main challenge*" (CT3)

After a positive experience with prototyping projects, CT1 transformed a contract-based relationship into a partnership early in their startup progress. The partner was firstly introduced to the CEO of CT1 via a personal relationship with the CTO. After the partner's competence was proof, two contracts were signed during the pre-startup and startup phases. The CEO satisfied with the speed of communication as well as delivery in general. The relationship between the CEO and the outsourcing team became closer and more personal when she visited the outsourcing team after three months of collaboration. After that, they exchanged personal contacts and had more social interactions. By learning the interest and business model from the outsource partner, the CEO sought to establish a more synergistic relationship by offering the firm to partner in some projects. Over time the relationship between the two firms became so strong that the CEO of CT1 issued shares of common stock to the outsourcing partner and added them in strategic and management meetings.

In summary, outsourcing relationship in startup phase often relies on previous positive experiences. Trust and personal relationship seems to influence the sustainability of the relationship. Requirements and expectations could go beyond contracts and relies on inter-personal governance.

### 3) Scaling phase

In the scaling phase, startups seek for break-even points, stable incomes, gaining strategic focus and professionalizing the organization of the team and market extending. Product development is under pressure of refactoring and extension. At this phase, the outsourcing relationship is transformed into an alliance in CT2 and the new outsourcing services are established to address the new technical tasks in CT5. In both case, there is a demand of scaling the services for a large amount of users (CT5) or enhancing the quality of services for new coming customers.

Under the increased amount of work, the CEO of CT2 invited the outsourcing team to take part of the startup, by offering some shares of the company. The CEO of CT2 had more than seven years working with Indian developers, which enables her to understand how they worked. She spent one month in

India to meet different teams and established a good relationship with the selected one. Early involvement in piloting was a good practice as mentioned by the CEO:

"*I have had experience with Indian developers earlier, and I like their attitudes … Cultural gaps hasn't been a problem. I have worked with them since 2008…*"(CT2)

She got to know the outsourcing team leader from a personal connection. Starting from doing rapid prototypes and mockups, the Indian team had gradually taken the main responsible for development and operation of the whole system. Over time, a deep alliance was formed between these two firms, where many maintenance and operational tasks were delivered by the Indian team at no cost. Due to this significant role, the CEO of CT2 gave a small part of company shares to the outsourcing partners, besides their periodic payment. In addition, the personal relationship was formed between the CEO of CT2 and the leader of the Indian team. They often exchanged opinions and advices related to their own business. Discussion about customers could also occur in weekends, and even they arranged a visit for the Indian leader to come to Norway.

In CT2, the CEO had a roadmap to develop three big features in the next 20 months. Besides the development, there was the need of maintaining current operating versions of the applications which were delivered to the customers. Establishment of partnership in CT2 enables risk management. Firstly, as the startup faced with new types of tasks when scaling up, many of them were not efficiently captured in an outsourcing agreement. Secondly, the startup team became more important when they were responsible for the quality of services delivered to an increasing number of customers. Besides, having the outsourcing team included in strategic meetings kept them informed timely with business decisions that impacts the product development.

After forming an alliance, the partnership had been going well, excepts for requests for more updates at the strategic level from CT2. A team member in the outsourcing team stated that:

"*At the moment, it might be a good idea to keep [Leader name] in the loop in future meeting, so we know better what is going there …*"(CT2)

In CT5, the company was under a pressure of resolving technical debts. During this time, the company moved the headquarter from Trondheim to London. CT5 adopted a software development team in London as a part of the company. They had done small outsourced tasks for the Trondheim's team. Having good experience with the team's performance and teamwork, the CEO of CT5 decided to acquire the team to join CT5. This caused the big change in organizational structure since CT5 moved to a new location. The former founder of CT5, who remains working in Trondheim, addresses the impact of the organizational change:

"*They have been working with [Feature name]. But recently I have little update on what is going there….*" (CT5)

In summary, startup's collaboration with external resource at the scaling phase often involve previous experience and a personal relationship. The shift to a form of partnership requires more commitment and engagement from both sides in sharing strategic and business information.

## V. DISCUSSIONS

### A. Startup experience with outsourcing

Reasons for outsourcing are similar to what we found in GSD literature, including cost saving, time to market, reaching talent pool [9, 26]. In many startups, it is important to have a prototype that illustrates technical and business capacity as soon as possible. In such case, startups are beneficial from outsourcing centers and R&D laboratories that are specialized on prototype development. Besides, many startups in early phases save administrative and legal costs by keeping minimum payrolls.

Failing factors in startup outsourcing are not surprising compared to what is known from GSD literature, i.e. lack of experience with GSD, poor delivered code quality; poor communication [9, 10], trust issue [33, 34], lack of mutual understanding [34, 35], and cultural and language barriers [15, 34]. A study about small-scale software projects showed that a focus on price when selecting outsourced partners could increase the likelihood of project failure [41]. By choosing the inappropriate partner, startups suffer from loosing time and potential market and business opportunities caused by the delay.

What is more particular to startup context is the lack of understanding and commitment to deal with the dynamic and uncertainties. As startup's business ideas are gradually refined and even pivoted during the time outsourcing contracts are carried on, unforeseen contingencies and contradictory requirements are important issues for outsourcing partners. Outsourcing teams that are used to outsourcing projects from established clients, might lack flexible methods and processes to deal with rapid changes in startups. Moreover, in many cases, CEOs and investors hired contracting teams without necessary knowledge and experience in software industry and particularly GSD.

We found evidence on positive outsourcing experience in cases of CT1b, CT1c, CT2b, CT3b, CT6. In the idea and pre-startup phases, outsourced projects are often small and contractual. All of the outsourced tasks are prototyping and small-scale. The positive experience in later phases of startups often derive from the previous successful relationships. For establishing a foundation for a long-term outsourcing relationship in a startup context, we identified the list of best practices for CEOs, including (1) communicating expectation and uncertainty with outsourced partners, (2) early investment on evaluating outsourcing services, (3) ensure an in-practice boundary spanner in play and (4) long-term plan for product roadmap.

### B. From contractual to partnership in startup context

In the early stages, our startups tried different outsourcing projects to develop a prototype to illustrate for their business domain. Our case sample are bootstrapping startups, which are lack of initial funding and resources. During the idea and pre-startup phase, the product development is limited to illustrating for technical or business viability. Outsourced projects at these phases are small-scale and experimental. Given that, startup founders who lack technical competence often choose outsourcing as a shortcut to a later stage of startups, where they can attract funding for proper product development. The contractual relationship started in this way often end up as a throw-away prototype, with lack of commitment and code quality.

In some other cases, startup founders looked for a sustainable strategy for product development, using their unique advantages, such as a personal relationship with a reliable outsourcing team, or successful collaboration previously. Quality insurance is an important task from the early time of the outsourcing relationship. The outsourcing delivery is expected to be evolutionary and turned into final products later. We observe that the shift from contractual outsourcing to partnership is a suitable path of startup evolution in case (parts of) core value is produced externally.

An important success factor for full-scale project is the role of a boundary spanner, who is familiar with language, cultural, mindset and working style of both startups and outsourced partners. In CT1, a regular communication channel has been established across geographical locations, including daily Skype meeting and an annual visit to the development site. Last but not least, the establishment of interpersonal relationship helps to overcome uncertainties, shortage of practices and processes in place.

Previous work defined five types of barrier to successfully manage a GSD project, which are geographical, temporal, cultural, work and process and organizational distances [10]. An outsourcing startup also needs to deal with the common challenges of geographical, temporal and cultural difference when choosing an outsourcing partner. Startups often start with a small team, lacking of proper development processes in place. Hence, in term of dealing with work and process issue, it is important to establish a practice of communication, delivery and quality assurance. When a contractual relationship evolves to a partnership, organizational distance is more of concern. The outsourcing partner would experience the lack of sharing identity, visions and timely updates with managerial decisions. This issue can be amplified by lower level of challenges. The GSD dimensions under the evolution of startup outsourcing relationship can be seen as in Figure 1.

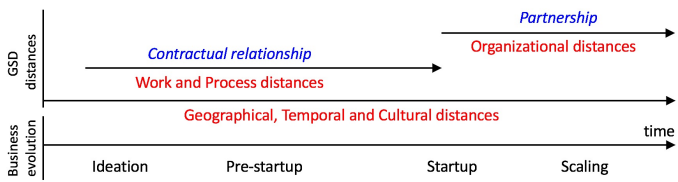

**Figure 1: Evolution of startup outsourcing**

## VI. CONCLUSIONS

Research on GSD has been substantially richer over the last decade. As there is no silver bullet for dealing with GSD projects, the best practices and evolution patterns need to be considered in a particular outsourcing setting. Considering the uniqueness of software startup as a SE context, this is one of the first studies investigate how software startups experience outsourcing relationship during different startup phases. As software startups are emerging as a multi-discipline research area, we are calling for investigation in different topics, such as business models, User Experience, team competence, startup growth patterns, etc [17]. The research on outsourcing practices and evolution in startup context contributes nicely to better understanding startup evolution and team competence strategies.

Our study makes several contributions to the software startup literature. Firstly, we portray successful and failure outsourcing experiences in software startup contexts. The experience was described according to stages in startup life cycle. Startups thus share common motivations and challenges when engaging in outsourcing projects. While the dynamic and uncertainty in product requirements introduce a new challenge in achieving expectation from outsourced partners, the experimental nature of early product development makes outsourcing a feasible option.

Secondly, we described and discussed common patterns in the evolution of outsourcing experience in startup contexts. A successful contractual relationship can evolve into a partnership in the startup and scaling phases. In such situation, the personal relationship is an important factor for going beyond contractual relationship and establishing the sustainability.

Future work could explore the outsourcing relationships in startups from USA or internal startups. Moreover, we plan to investigate the issues of intellectual property when outsourcing in the next step. With a wider viewpoint on startups, we will present a guideline with best practices for outsourcing startup.